\documentclass[aps,prd,nofootinbib,showpacs,amssymb,twocolumn]{revtex4}
\usepackage{amsmath,amsfonts}
\usepackage{epsfig,subfigure}
\usepackage{ifpdf}
\usepackage[bookmarks=false]{hyperref} %
\usepackage{amsmath}
\usepackage{graphicx}
\usepackage{color}
\usepackage{natbib}
\usepackage{multirow}

\usepackage[normalem]{ulem} %

\newcommand{\be}{\begin{equation}}
\newcommand{\ee}{\end{equation}}
\newcommand{\bea}{\begin{eqnarray}}
\newcommand{\eea}{\end{eqnarray}}

\newcommand\ILE{ILE}

\newcommand\hidetosubmit[1]{}
\renewcommand\hidetosubmit[1]{#1}
\newcommand\optional[1]{}
\newcommand\ForInternalReference[1]{}

\newcommand\unit[1]{\, {\rm #1}}

\newcommand\Y[1]{Y^{(#1)}}

\newcommand\qmstateproduct[2]{\left<#1|#2\right>}

\begin{document}

\title{An architecture for efficient  gravitational wave parameter estimation with multimodal linear  surrogate models}
\date{\today}
\author{Richard O'Shaughnessy}
\email{oshaughn@mail.rit.edu}
\affiliation{Center for Computational Relativity and Gravitation, Rochester Institute of Technology, Rochester, NY 14623, USA}

\author{Jonathan Blackman}
\affiliation{TAPIR, Walter Burke Institute for Theoretical Physics, Mailcode 350-17, California Institute of Technology, Pasadena, CA 91125, USA}

\author{Scott E. Field}
\affiliation{Mathematics Department, University of Massachusetts Dartmouth, Dartmouth, MA 02747, USA}
\affiliation{Cornell Center for Astrophysics and Planetary Science, Cornell University, Ithaca, New York 14853, USA}

\begin{abstract}
The recent direct observation of gravitational waves has further emphasized the desire for fast, low-cost, and accurate methods to infer the parameters of gravitational wave sources. 
Due to expense in waveform generation and data handling, the cost of evaluating the likelihood function limits the computational performance of these calculations.
Building on  recently developed surrogate models and a novel parameter estimation pipeline, we show how to  quickly
generate the likelihood function as
an analytic, closed-form expression.  Using a 
straightforward
variant of a production-scale parameter
estimation code, we demonstrate our method using 
surrogate models of effective-one-body and numerical relativity waveforms.   Our study is the first time these models
have been used for parameter estimation and one of the first ever parameter estimation calculations with
multi-modal numerical relativity waveforms, which include all $\ell \leq 4$ modes.
Our  grid-free method enables rapid parameter estimation for any waveform with
a suitable reduced-order model.
The methods described in this paper may also find use in other data analysis 
studies, such as vetting coincident events or the computation of the coalescing-compact-binary detection statistic.
\end{abstract}
\maketitle

\section{Introduction}

On September 14, 2015, at 09:50:45 UTC, 
the Laser Interferometer Gravitational-Wave Observatory (LIGO) made the first direct 
observation of a gravitational-wave signal from two coalescing black hole binaries \cite{DiscoveryPaper}. By
systematically comparing the signal against approximations to the solutions 
of Einstein's equations, the properties of
the coalescing black hole binary  were inferred \cite{PEPaper}. 
Similar binary black hole systems will be detected in coming
years \cite{RatesPaper,LIGO-O1-BBH}, at a rate of up to one per day by advanced LIGO and Virgo.  These discoveries will revolutionize our understanding of
astrophysics (see, e.g., 
Refs.~\cite{AstroPaper,2016Natur.534..512B,2016ApJ...824L...8R,2016PhRvL.116t1301B,2016MNRAS.460.3545D} and references therein) and provide tests of gravitational theory to unprecedented accuracies 
in the regime of strong-field dynamics
with relativistic velocities~\cite{TheLIGOScientific:2016src,LIGO-O1-BBH}.

The  vigorous pace of discovery, combined with the tantalizing opportunities afforded by low-latency and coordinated
multimessenger observations \cite{2016LRR....19....1A}, demand equally rapid inference: LIGO and its electromagnetic partners should prepare to
reliably reconstruct the source parameters of coalescing binaries as fast as possible \cite{LIGO-whitepaper-2015}.  
Especially when using the best-available waveform models, these calculations can be very costly; see, e.g., \cite{SEOBv3Paper}.

Several strategies have been
developed to reduce the 
computational cost of parameter estimation \cite{gwastro-pe-Brandon-STF2,gwastro-PE-AlternativeArchitectures,gw-astro-ReducedOrderQuadraturePE-TiglioEtAl2014,2016PhRvD..94d4031S,2017arXiv170302062V}.  
Approaches that have appeared in the literature include
generating the approximate solutions more quickly \cite{gwastro-mergers-PE-ReducedOrder-2013,2014PhRvX...4c1006F,2013PhRvD..87l2002S,2013PhRvD..87d4008C,gwastro-mergers-IMRPhenomP,gwastro-SpinTaylorF2-2013}; 
interpolating some combination
of the waveform or likelihood \cite{gwastro-approx-ROMNR-Blackman2015,2013PhRvD..87d4008C,2013PhRvD..87l2002S,2014CQGra..31s5010P,2014PhRvD..90d4074S,gwastro-PE-AlternativeArchitectures,cole2014likelihood,2012MNRAS.421..169G};  or 
adopting a sparse representation to reduce the computational cost of data
handling \cite{antil2013two,gwastro-mergers-PE-ReducedOrder-2013,2016PhRvD..94d4031S,gw-astro-ReducedOrderQuadraturePE-TiglioEtAl2014,gwastro-PE-AlternativeArchitectures}.   %
 Some methods, however, achieve rapid turnaround through simplifying approximations.  

Two rapid strategies eschew significant approximation: reduced-order models (ROMs), a term we shall use interchangeably with surrogate models, and refactored likelihoods. 

A surrogate
model provides an efficient and highly accurate representation for the gravitational wave strain.
Surrogate models  have been applied to reproduce the radiation from
complicated sources, including long duration signals~\cite{2014PhRvX...4c1006F,2014CQGra..31s5010P}, arbitrarily many
harmonic modes~\cite{2014PhRvX...4c1006F,gwastro-approx-ROMNR-Blackman2015}, spinning binary systems~\cite{2014CQGra..31s5010P,purrer2016frequency}, precessing binary systems~\cite{blackman:4d2s,2016PhRvD..94d4031S}, and neutron star models with tidal effects~\cite{lackey2016effective}.
Moreover, as we describe in this paper, 
calculations that arise naturally in parameter estimation studies
can be 
expressed in terms of simple, precomputed quantities
constructed from the reduced-order representation.
The result is a dramatic reduction in
the number and complexity of operations needed to carry out 
gravitational-wave inference.

Similarly, Pankow et
al. \cite{gwastro-PE-AlternativeArchitectures} (henceforth \ILE, a shorthand for ``integrate likelihood over the extrinsic parameters'') expressed the gravitational-wave
strain using the natural basis
provided by a spin-weighted spherical harmonic decomposition of the waves.  By almost
eliminating overhead from data handling (e.g., the cost of performing Fourier transforms and inner products needed to evaluate ${\cal L}$), this
representation allows for
rapid likelihood evaluations, enabling direct Monte Carlo integration over all ``fast''
 variables (e.g., extrinsic parameters corresponding to the spacetime location and orientation of the binary, which
   leave the binary's intrinsic dynamics unchanged).
The ILE framework was recently applied in
\cite{NRPaper}, to directly compare GW150914
against 
numerical simulations of Einstein's
equations, without any intermediate approximation.

In this work, we demonstrate that these two approaches can and should be naturally unified, dramatically enhancing 
overall performance. 
This combination increases the performance of \ILE{} by removing the need for a brute-forced grid-based exploration of the
  intrinsic parameter space, which can be a source of error.
Additionally, we present the first parameter estimation results using multimodal numerical relativity
surrogates.

This paper is organized as follows.  In Section \ref{sec:Methods} we 
introduce
the likelihood
calculation (Section \ref{sec:sub:PEBackground}), 
describe how to refactor the log-likelihood
for efficient use of reduced-order 
gravitational-wave models (Section \ref{sec:sub:Refactor}), 
and 
implement
our procedure as a simple
extension of an existing, production scale parameter estimation pipeline
by interfacing this pipeline with low-level
surrogate data-access tools (Section \ref{sec:sub:Implementation}).
Section \ref{sec:Results}
demonstrates the method using end-to-end comparisons with the traditional \ILE{} framework. For simplicity and to facilitate illustrations and
comparisons, we emphasize 
examples
using a previously-reported and widely available nonspinning, comparable-mass
effective-one-body (EOB) surrogate~\cite{2014PhRvX...4c1006F}\footnote{\label{foot:GWS}This surrogate model, which is distributed with the \texttt{gwsurrogate} package~\cite{gwsurrogate}, was built for the EOB model described in Ref.~\cite{pan2011inspiral} and implemented in the routine EOBNRv2 as part of the publicly available LIGO Analysis Library (LAL) Suite. The git hash 
59c12886b026c863397f191e6c2ca69ef3498616 (available, e.g., at \texttt{https://github.com/lscsoft/lalsuite})} provides the exact code snapshot of LAL at the time the surrogate was built.
and a nonspinning numerical relativity surrogate model~\cite{gwastro-approx-ROMNR-Blackman2015} 
including
$77$ harmonic modes up to $\ell=8$ and trained on a mass ratio interval of $q = m_1 / m_2 \in [1,10]$, where $m_1$ and $m_2$ are the binary's component masses.

\section{Methods}
\label{sec:Methods}

\subsection{Inference by (Monte Carlo) integration}
\label{sec:sub:PEBackground}

\subsubsection{Preliminaries}

Given a value of the intrinsic parameters $\lambda$ (eight parameters characterizing the two masses and spin vectors)
and extrinsic parameters $\theta$ (four spacetime coordinates for the coalescence
event; three Euler
angles for the binary's orientation relative to the Earth),  we can predict the response $h_k$ of 
LIGO's two operational instruments, denoted as  $k=\{1,2\}$, to an impinging
gravitational wave signal.
Assuming a Gaussian, stationary noise model, we can evaluate the log-likelihood
\begin{align} \label{eq:loglikelihood}
\ln & {\cal L}(\lambda, \theta) = \nonumber \\
& -\frac{1}{2}\sum_k \qmstateproduct{h_k(\lambda,\theta)-d_k}{h_k(\lambda,\theta)-d_k}_k  
 - \qmstateproduct{d_k}{d_k}_k
\end{align}
of LIGO's network of observatories having recorded a gravitational wave signal.
Except for the overall normalization constant,
and omitting calibration uncertainty,
our expression~\eqref{eq:loglikelihood}
agrees with Eq. (1) in \cite{PEPaper}.
Here $d_k$ is the
detector data in instrument $k$, 
\begin{align*}
\qmstateproduct{a}{b}_k \equiv \int_{-\infty}^{\infty} 2 df \frac{\tilde{a}(f)^*\tilde{b}(f)}{S_{n,k}(|f|)} \,,
\end{align*}
is a 
noise-weighted inner product implied by the $k$th-detector's noise power spectrum
$S_{n,k}(f)$, $\tilde{a}(f)$ is the Fourier transform of $a(t)$,
$\tilde{a}(f)^*$ denotes complex conjugation of $\tilde{a}(f)$,
and $f$ is frequency;  see, e.g., \cite{gwastro-PE-AlternativeArchitectures} for more details. 
In practice, and as discussed in the
next section, we adopt a low-frequency cutoff $f_{\rm low}$
such that all inner products are modified to
\begin{eqnarray}
\qmstateproduct{a}{b}_k\equiv 2 \int_{|f|>f_{\rm low}}  df \frac{\tilde{a}(f)^*\tilde{b}(f)}{S_{n,k}(|f|)} \,.
\end{eqnarray}

A key task of any parameter estimation study is to 
compute
the joint posterior probability of $\lambda,\theta$
\begin{eqnarray}
\label{eq:Posterior:General}
p_{\rm post}(\lambda,\theta) = \frac{{\cal L}(\lambda,\theta) p(\theta) p(\lambda)}{ \int d\lambda d\theta {\cal
  L}(\lambda,\theta) p(\lambda)p(\theta)} \; ,
\end{eqnarray}
which follows from Bayes' theorem.
Here $p(\theta)$ and $p(\lambda)$ are priors on the (independent) variables $\theta,\lambda$.\footnote{For simplicity,
  we assume all 
binary black hole systems are equally likely anywhere in the universe, at any orientation relative to the detector. Future direct observations may favor a
  correlated distribution, including  
the formation of more massive black holes at larger redshift \cite{2016Natur.534..512B}.
}

\subsubsection{Fast and slow intrinsic parameters}

Following \cite{gwastro-PE-AlternativeArchitectures}, we 
partition the intrinsic parameter $\lambda$ into ``fast'' and ``slow''
parameters denoted by $\lambda_f$ and $\lambda_s$, respectively.
In principle, this  division depends entirely on the computational cost of waveform generation. 
``Fast" parameters are those for which new waveform evaluations can be quickly generated from existing ones as the value of $\lambda_f$ changes. Typically, this is accomplished by an explicit, closed-form expression. In the original ILE study~\cite{gwastro-PE-AlternativeArchitectures}, the fast (slow) parameters were the extrinsic (intrinsic) parameters. By contrast, for this paper, and as we show for any other ILE-based investigation using linear surrogates that represent the scale-free general relativity solution,
the only ``slow'' parameter is the binary system's total mass $M = m_1 + m_2$.

Having split the intrinsic parameters into a fast and slow set, 
we shall now view the likelihood function as ${\cal L}(\lambda_f, \lambda_s, \theta)$. For a fixed
value of $\lambda_s$, integration over
all fast parameters %
leads to an intermediate result\footnote{In general the prior in $\lambda_s,\lambda_f$ will not be separable: the range of allowed mass ratios will depend on total mass, for example.}:
\begin{eqnarray}
\label{eq:Marginalize:GeneralIdea}
{\cal L}_{\rm marg,s}(\lambda_s) \equiv \int {\cal L}(\lambda,\theta)p(\theta) p(\lambda_s,\lambda_f) d\theta d\lambda_f\ .
\end{eqnarray}
Note that unlike in the original ILE framework, we have explicitly retained the prior $p(\lambda_s,\lambda_f)$ 
in this expression.
For the \ILE{} study \cite{gwastro-PE-AlternativeArchitectures},  $\lambda_s$  included all intrinsic parameters, with
$\lambda_f$ being empty.  
 In that work, a function proportional to ${\cal L}_{\rm marg,s}(\lambda = \lambda_s)$ was evaluated on a grid;
interpolated, fitted, or otherwise approximated; and hence used to 
generate the posterior  as a function of $\lambda_s$
\begin{eqnarray}
p_{\rm post}(\lambda_s) =  \frac{{\cal L}_{\rm marg,s }(\lambda_s)  }{ \int d\lambda_s  {\cal
  L}_{\rm marg,s}(\lambda_s )} \; ,
\end{eqnarray}
which follows by integrating Eq. (\ref{eq:Posterior:General}) over $\theta$ and $\lambda_f$.\footnote{Note that in this expression (and
in contrast to the notation in \cite{gwastro-PE-AlternativeArchitectures}), our expression for ${\cal L}_{\rm marg}$
includes the prior over $\lambda$, allowing us to employ the same expression for the posterior to describe the method
used in this work and in \cite{gwastro-PE-AlternativeArchitectures}. }
The denominator of this quantity, the (Bayesian) evidence for our model, can be used to assess how well our model fits the data.
The ILE grid-based design was intended
to minimize the severe computational cost
of evaluating waveforms at different values of $\lambda_s$.
In Sec.~\ref{sec:sub:Refactor}, we show how linear surrogate models remove any need for expensive, high-dimensional grids.

If the 
integral appearing in Eq.~\eqref{eq:Marginalize:GeneralIdea} is
performed by direct Monte Carlo integration, and this 
computation
is repeated on a dense and uniform grid in
$\lambda_s$ (here, $\lambda_s=M$, the total binary mass), the posterior may 
be estimated using a fit-free method.
ILE used this same method -- henceforth denoted ILEMC --  to infer posterior distributions in $\theta$.  Assuming we have $N$ random samples 
$\{\theta_q,\lambda_{f,q}\}_{q=1}^N$
drawn from 
a sampling distribution $p_s(\lambda_f,\theta; \lambda_s)$ at fixed value of $\lambda_s$, then 
the numerical approximation $\hat{{\cal L}}_{\rm marg,s}(\lambda_s)$
to the true marginalized likelihood ${\cal L}_{\rm marg,s}(\lambda_s)$ 
computed by Monte Carlo integration is:
\begin{subequations}
\label{eq:MonteCarlo}
\begin{eqnarray}
\hat{{\cal L}}_{\rm marg,s}(\lambda_s) &=&  \frac{1}{N} \sum_{q=1}^N w_q (\lambda_s) \,, \\
w_q (\lambda_s) &=& \frac{{\cal L}(\lambda_s,\lambda_{f,q},\theta_q) p(\theta_q) p(\lambda_{s},\lambda_{f,q})}{p_s(\lambda_{f,q},\theta_q; \lambda_s) } \,.
\end{eqnarray}
\end{subequations}
We repeat this process for a uniform grid in $\lambda_s$, using
the same number of samples, N, each time.  As a result and in particular, we can estimate the true one-dimensional cumulative
distribution $P(< x)$ by the numerical approximation $\hat{P}(<x)$:
\begin{eqnarray}
\label{eq:MCPosterior:1d}
P(< x) \approx \hat{P}(<x) = \frac{\sum_g \Theta(x-x_g) w_g}{\sum_g w_g} \,,
\end{eqnarray}
i.e., by a Monte Carlo integral over the interval $<x$.
Here $x$ can be any parameter in $\lambda$ or $\theta$, 
$g$ indexes the Monte Carlo samples over the union of all 
of the 
values of $\lambda_s$, $x_g$ refers to the value of parameter $x$ for the $g$th sample,
and $\Theta(x)$ is the Heaviside step function.
Using higher-dimensional weighted density estimates (e.g., kernel density estimators or weighted histograms), we can
likewise estimate the joint posterior distribution in any set of dimensions, with sufficiently dense sampling.

\subsection{Refactored likelihood for linear surrogates}
\label{sec:sub:Refactor}

\subsubsection{Surrogate-enabled ILE}

A complex gravitational-wave strain
\begin{align} \label{eq:strain}
h(t,\vartheta,\phi;\lambda) =  h_+(t,\vartheta,\phi;\lambda) - 
                                i h_\times (t,\vartheta,\phi;\lambda) \, ,
\end{align}
can be expressed in terms of its two fundamental polarizations $h_+$ and $h_\times$.
Here, $t$ denotes time, $\vartheta$ and $\phi$ are the polar and azimuthal angles
for the direction of gravitational wave propagation away from the source. 
The complex gravitational-wave strain can be written in terms of
spin-weighted spherical harmonics $\Y{-2}_{\ell m} \left(\vartheta, \phi \right)$ as 
\begin{align} \label{eq:strain_mode}
h(t,\vartheta,\phi;\lambda) = 
\sum_{\ell=2}^{\infty} \sum_{m=-\ell}^{\ell} \frac{D_{\rm ref}}{D} h^{\ell m}(t;\lambda) \Y{-2}_{\ell m} \left(\vartheta, \phi \right) \, ,
\end{align}
where the sum includes all harmonic modes $h^{\ell m}(t;\pmb{\lambda})$ made available by the model;  where
$D_{\rm ref}$ is a fiducial reference distance; and where $D$, the luminosity distance to the  source, is one of the
extrinsic parameters.  

Following a standard ROM prescription, we assume access to a {\em linear} surrogate model for 
each harmonic mode
\begin{eqnarray} \label{eq:linear_surrogate}
h_{\ell m}(t; \lambda) = \frac{M}{D_{\rm ref}} c_{\ell m, \alpha}(\lambda_f) W_\alpha(t/M) \,,
\end{eqnarray}
associated with some  fiducial distance $D_{\rm ref}$, which can be expressed as a {\em linear} expansion
in a set of reduced basis functions $W_\alpha$.
Different surrogate modeling techniques 
prescribe different approaches for the coefficients $c_{\ell m, \alpha}$ -- these details need not concern us
here. Crucially, this decomposition naturally identifies the ``fast'' and ``slow'' intrinsic parameters.
  Since the time and total mass are coupled through the basis functions, $W_\alpha(t/M)$, 
  we are unable to ``pull" $M$ out of the inner products (cf.~Eq.~\eqref{eq:QuantitiesViaSurrogate}); computing the 
  likelihood for new values of $M$ will require the computation of many slow overlap integrals. 
  Conversely, as the remaining intrinsic parameters only enter through the expansion
  coefficients, $c_{\ell m, \alpha}(\lambda_f)$, computing the likelihood for new values of 
  $\lambda_f$ is accomplished with fast evaluations of these coefficients (cf.~Eq.~\eqref{eq:QuantitiesViaSurrogate}).

This approach requires that
the coefficients are given by a known, closed-form expression 
and the model's temporal and parametric dependence has an affine factorization of Eq.~\eqref{eq:linear_surrogate}. 
In particular, this restriction precludes surrogate models whose temporal dependence has
a {\em non-linear} relationship to the basis (say, by an amplitude and phase decomposition). As described later on, 
we do not believe this to be any real restriction in practice since all surrogates can be brought into the form~\eqref{eq:linear_surrogate}. Indeed, the numerical relativity surrogate model used in this paper~\cite{gwastro-approx-ROMNR-Blackman2015}, expressed as two independent linear expansions of the amplitude and
phase of each mode $h_{\ell m}(t,\lambda) $, was originally in a format incompatible with the representation~\eqref{eq:linear_surrogate}.

Following \citet{gwastro-PE-AlternativeArchitectures}, we substitute expression~\eqref{eq:strain_mode} 
for $h_{\ell m}$ into the expression $h_k(t-t_k) =F_{+,k} h_+(t-t_k) +
  F_{\times,k}h_\times(t-t_k)$ for the detector response $h_k$,
where $t_k=t_c - \vec{x}_k \cdot \hat{n}$ is the arrival time at the $k$th detector (at position $\vec{x}_k$)
for a plane wave propagating along $\hat{n}$ 
and $t_c$ is the time of coalescence~\cite{gwastro-PE-AlternativeArchitectures}.
We then substitute these expressions for $h_k$ into the likelihood function~\eqref{eq:loglikelihood}
thereby generating~\cite{gwastro-PE-AlternativeArchitectures}
\begin{widetext}
\begin{align}
\ln {\cal L}(\lambda, \theta) 
&= (D_{\rm ref}/D) \text{Re} \sum_k \sum_{\ell m}(F_k \Y{-2}_{\ell m})^* Q_{k,lm}(\lambda,t_k)\nonumber \\
&   -\frac{(D_{\rm ref}/D)^2}{4}\sum_k \sum_{\ell m \ell' m'}
\left[
{
|F_k|^2 [\Y{-2}_{\ell m}]^*\Y{-2}_{\ell'm'} U_{k,\ell m,\ell' m'}(\lambda)
}
 {
+  \text{Re} \left( F_k^2 \Y{-2}_{\ell m} \Y{-2}_{\ell'm'} V_{k,\ell m,\ell'm'} \right)
}
\right]
\label{eq:def:lnL:Decomposed}
\end{align}
\end{widetext}
where 
where $F_k = F_{+,k} - i F_{\times,k}$ are the
complex-valued detector
response functions of the $k$th detector \cite{gwastro-PE-AlternativeArchitectures} and
the quantities $Q,U,V$ depend on $h$ and the data as
\begin{subequations}
\label{eq:QUV}
\begin{align}
Q_{k,\ell m}(\lambda,t_k)
&= 2 \int_{|f|>f_{\rm low}} \frac{df}{S_{n,k}(|f|)} e^{2\pi i f t_k} \tilde{h}_{\ell m}^*(\lambda;f) \tilde{d}(f)\ , \\
{ U_{k,\ell m,\ell' m'}(\lambda)}& = \qmstateproduct{h_{\ell m}}{h_{\ell'm'}}_k\ , \\
V_{k,\ell m,\ell' m'}(\lambda)& = \qmstateproduct{h_{\ell m}^*}{h_{\ell'm'}}_k  \ .
\end{align}
\end{subequations}
Finally, substituting~\eqref{eq:linear_surrogate} into \eqref{eq:QUV}
while fixing the value of $M$,
we find that all three parameter-dependent functions can be expressed in terms of the
surrogate interpolating functions, $c_{\ell m,\alpha}(\lambda)$, and 
correlations between the reduced basis functions and data:
\begin{subequations}
\label{eq:QuantitiesViaSurrogate}
\begin{eqnarray}
Q_{k,\ell m}(\lambda,t_k) = c_{\ell m,\alpha}(\lambda) \qmstateproduct{T_{t_k} W_\alpha}{d_k}_k \,, \\
U_{k,\ell m,\ell m'}(\lambda) = c^*_{\ell m,\alpha}(\lambda)   \qmstateproduct{W_\alpha}{W_\beta}_k c_{\ell'm',\beta}(\lambda)  \,, \\
V_{k,\ell m,\ell m'}(\lambda) = c_{\ell m,\alpha}(\lambda)   \qmstateproduct{W^*_\alpha}{W_\beta}_k c_{\ell'm',\beta}(\lambda)  \,.
\end{eqnarray}
\end{subequations}
Here $T_\tau$ is a time-translation operator $(T_\tau f)(t)=f(t-\tau)$. 
In other words, after a handful of inner product evaluations that can be computed once and for all, the likelihood can be subsequently evaluated for all
intrinsic parameters $\lambda$ \emph{analytically}, as $c(\lambda)$ are known analytic expressions (made directly available by the surrogate model) and $Q,U,V$ are easily tabulated arrays. 

\subsubsection{Discussion and further compression}
\label{sec:compression}

Equation~\eqref{eq:QuantitiesViaSurrogate} 
needs to be evaluated for each basis function
in the surrogate 
model.
Current multimodal surrogates (including the surrogates employed here) treat each mode independently, so the total number of basis functions 
grows with the number of modes 
as well as
the number of basis functions per mode.
For example, in some of our parameter estimation studies we use a non-spinning NR surrogate that contains $77$ modes and about $25$ basis per mode, implying
an enormous number of inner products~\eqref{eq:QuantitiesViaSurrogate}.

The basis size can be reduced three ways. Our procedure follows a combination of the first two observations described
 below; we hope to explore the final approach (which requires building a new surrogate model)  in future work.

\begin{figure}
\includegraphics[width=\columnwidth]{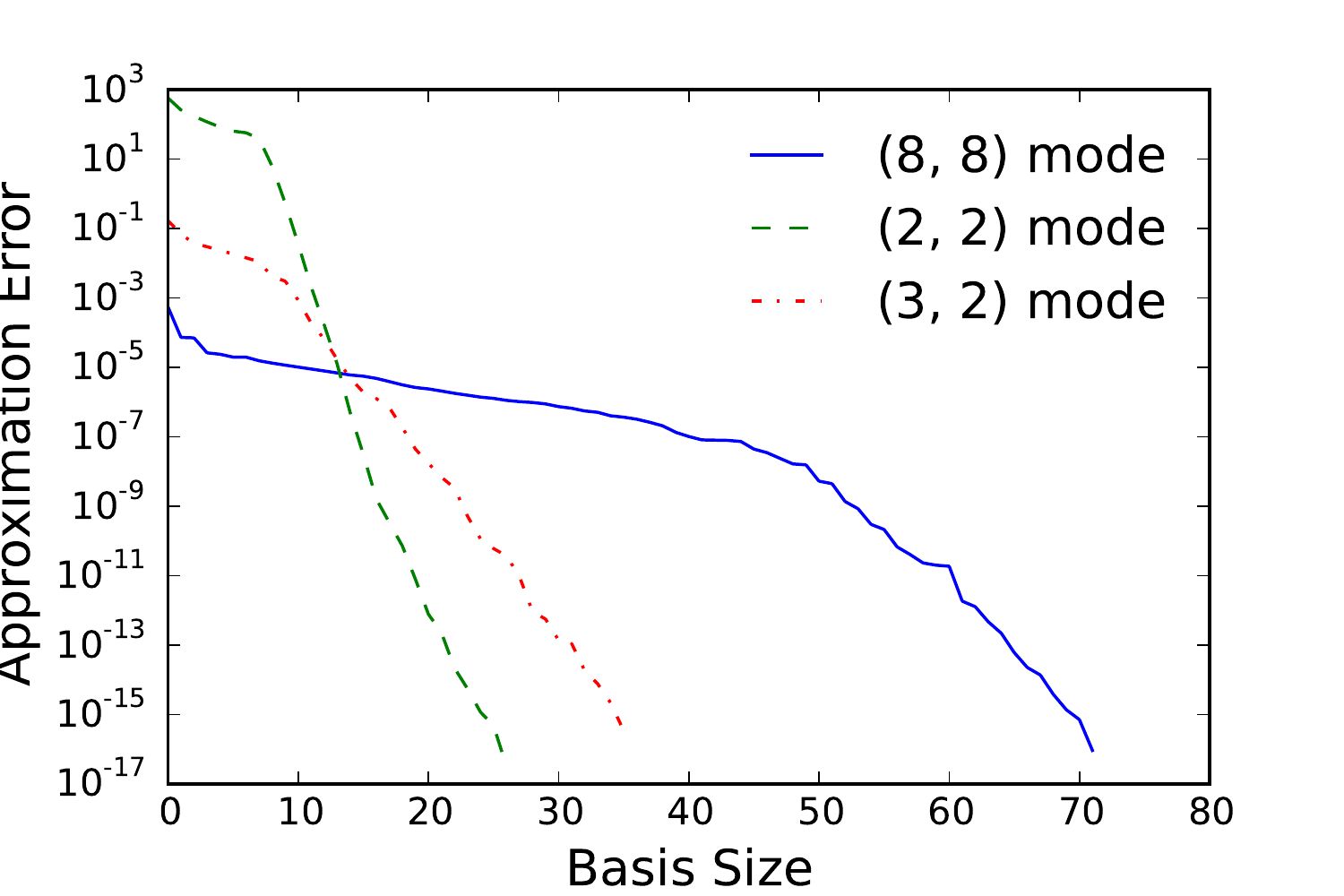}
\caption{\label{fig:basis_convergences}\textbf{Convergence of the NR surrogate basis}: Convergence of the reduced basis approximation (i.e. for an optimal set of expansion coefficients $c_{\ell m, \alpha}(\lambda)$ found through orthogonal projection)
of the linear NR surrogate model, $h^{\ell m}_\mathrm{S,1}(t,\lambda)$, for 
three representative harmonic modes. 
The $L^2$-type errors, computed from the formula $\max_{\lambda} \| h^{\ell m}_\mathrm{S,1}(\cdot; \lambda) - h^{\ell m}_\mathrm{S,2}(\cdot; \lambda)\|$, are measured with respect to the original {\em non-linear} surrogate, $h^{\ell m}_\mathrm{S,2}(t;\lambda)$, built in Ref.~\cite{gwastro-approx-ROMNR-Blackman2015}. These {\em un-normalized} errors should be compared on a relative scale. In
particular, we see that, as expected, the (2,2)-mode's accuracy dominates the overall error budget. 
By contrast to the quadrupole mode, the (8,8) mode is relatively unimportant as evidenced by six orders of magnitude difference with the $(2,2)$ mode. (Using zero basis functions is equivalent to neglecting the mode altogether.) Our linear surrogate truncates the basis size at $22$ regardless of harmonic mode index, and the ILE pipeline makes additional mode-specific basis reductions based on the truncation's impact to the overall surrogate model accuracy (see Sec.~\ref{sec:compression}).
}
\end{figure}

First, we can eliminate superfluous modes from our expansion; for example, the $(l,m)=(8,8)$ mode is rarely practically
relevant (i.e., $\qmstateproduct{h_{88}}{h_{88}} \ll \qmstateproduct{h}{h}$; cf. the caption of Fig.~\ref{fig:basis_convergences}).  Our ILE implementation automatically eliminates modes which are unlikely to be relevant, based on a reference
set of parameters provided by the gravitational wave search.

Second, we can easily reduce the number of basis elements
needed per mode.
As shown in Fig.~\ref{fig:basis_convergences}, surrogate errors typically converge
exponentially and employ an excessively accurate basis (often with overlap errors around $10^{-10}$) for parameter estimation purposes. So the basis size can be reduced by as much as a factor of 2 in our case
without a significant loss in accuracy.\footnote{To assess just how excessively accurate these surrogate model
  constructions can be, we compare the typical accuracy shown in Fig. (\ref{fig:basis_convergences}) to the largest mismatch error that could introduce a significant
  deviation into a parameter estimation posterior: an overlap error of order $0.1/\rho^2 \simeq 10^{-5}(\rho/100)^2$. This nominal overlap error
  is also often significantly smaller than other systematic effects associated with intrinsic NR error due to finite
  resolution and extraction, as well as to effects associated with the neglect of higher-order modes.} 
To be concrete, 
Eq.~\eqref{eq:QuantitiesViaSurrogate} can be
expressed in terms of orthogonal basis functions, $e_\alpha(t)$, which are related to the 
basis functions, $W_\alpha(t)$, by
a linear transformation, $W_\alpha(t) = {\cal V}_{\alpha \beta} e_\beta(t)$.
The transformation matrix ${\cal V}_{\alpha \beta}$ is a necessary part of the surrogate building process~\cite{2014PhRvX...4c1006F} and is readily available for use by the ILE codes. In the orthogonal representation, the
basis elements are ordered by significance; we can therefore dramatically reduce the number of basis elements needed, by
adopting a basis size suitable to the comparatively lower accuracy needed for our calculations.   
For simplicity and modularity, however, our current implementation uses Eq. (\ref{eq:QuantitiesViaSurrogate}) directly,
without additional refactoring. Instead, our current implementation can drop higher-order (orthogonal) basis functions before the
computation of $W_\alpha$ and hence Eq.~\eqref{eq:QuantitiesViaSurrogate}.

Finally, we can reduce the overall set of $W_{\alpha}$ by using another surrogate construction procedure, employing the
same basis set for all the modes.  This straightforward second-generation surrogate should require 
fewer basis coefficients, particularly since modes 
with the same harmonic index $m$ share similar frequency content at early times.

\subsubsection{Implementation considerations}

Our approach requires the surrogate model
to be expressed as a linear combination of basis elements.
However, not all surrogates have this form. For example, some models are built to separately reconstruct the amplitude and
phase of $h_{\ell m}$ see, e.g., \cite{2014CQGra..31s5010P,gwastro-approx-ROMNR-Blackman2015}. 
That said, any surrogate can be used to train a secondary ``surrogate-of-a-surrogate"
that has the necessary form given by Eq.~\eqref{eq:linear_surrogate}.
We have found that building a secondary surrogate is significantly easier than the original surrogate since (i) the
waveform training data is already aligned [i.e., the surrogate-builder does not need to duplicate the effort needed to
  establish a consistent definition of the event time]\footnote{In our case, the initial surrogate was aligned so that the maximum of $\sum_{\ell m}|h_{\ell m}(t)|^2$ occurs at $t=0$; see Eq.~(2) in Ref.~\cite{gwastro-approx-ROMNR-Blackman2015}.} and (ii) arbitrarily many waveform evaluations can be supplied by the primary
surrogate.  Moreover,
several modern surrogates are already expressed in 
as a linear combination of basis to enable their use in reduced-order-quadrature
methods \cite{2016PhRvD..94d4031S}.   Our method can therefore be applied to all available surrogates without loss of generality.
As described in \cite{2014PhRvX...4c1006F}, the  \texttt{gwsurrogate}
package~\cite{gwsurrogate} provides an interface to generic surrogates for gravitational wave
radiation from coalescing binaries.  As part of this paper, we have extended gwsurrogate's API to allow for a convenient 
interface with low level surrogate waveform data as needed by the surrogate-enabled ILE pipeline.  In this work we will demonstrate our method 
using one of the surrogates provided with it: a zero-spin, equal-mass
surrogate   tuned to the nonspinning effective-one-body model~\cite{pan2011inspiral}. For each total mass
$M$,  we can extract the basis functions  $W_\alpha$ and construct the inner products appearing in
Eq. (\ref{eq:QuantitiesViaSurrogate}), once and for all.   To evaluate the likelihood any mass ratio $q$ and extrinsic
parameters $\theta$, we use the   \ILE{} likelihood [Eq. (\ref{eq:def:lnL:Decomposed})], where 
 $Q,U,V$ are evaluated using Eq. (\ref{eq:QuantitiesViaSurrogate}) and the coefficients $c_{\ell m,\alpha}(q)$ are provided by our surrogate model.

\subsection{Two methods to infer parameters}
\label{sec:sub:Implementation}

In Sec.~\ref{sec:Results} we will directly compare two ILE-type
approaches
to infer parameters: traditional ILE~\cite{gwastro-PE-AlternativeArchitectures}
and its extension developed here.
While both methods have been presented for generic binary 
black hole systems, the parameter estimation results of Sec.~\ref{sec:Results}
are for non-spinning binaries. And so, for concreteness, 
we briefly summarize these methods when specialized to such systems.

\noindent \textbf{Traditional ILE.} 
We use ILE to (i) evaluate ${\cal L}_{\rm marg,alt}(m_1,m_2) = \int d\theta
p(\theta){\cal L}(\lambda,\theta)$ by direct Monte Carlo integration,
(ii) fit this function, as $\hat{\cal L}_{\rm marg,alt}$~\footnote{To ensure robust results, we have employed both  low-order polynomial and Gaussian process fits; our
  results do not change significantly (i.e., the average difference $\int_{x_1}^{x_2} dx |P_A(<x) -
    P_B(<x)|/(x_2-x_1)$ between  two cumulative distribution function estimates $P_A,P_B$ is a few percent over the
    interval $x_1,x_2$ shown in our figures). },
and then (iii) integrate  $p_{\rm post}(\lambda)= \hat{{\cal
  L}}_{\rm marg,alt} p(\lambda)/\int d \lambda  \hat{{\cal
  L}}_{\rm marg,alt} p(\lambda)$ to evaluate the posterior. 

In contrast to ${\cal L}_{\rm marg, s}(\lambda_s)$ 
given by Eq.~\eqref{eq:Marginalize:GeneralIdea}, which is a function only of the total binary mass
$\lambda_s=M$ and which in this work is used only for normalization, the function ${\cal L}_{\rm marg,alt}(\lambda)$
depends on all intrinsic binary parameters $\lambda$.

 We perform the 
integration carried out in the third step via Monte Carlo, using a uniform prior density in
$(m_1,m_2)$ 
such that each component mass is greater than $1 M_\odot$ and the total mass is less than $200
 M_\odot$.   
The prior's boundary is defined by a right triangle with verticies at $(1M_\odot,1M_\odot)$, $(199 M_\odot,1M_\odot)$, and $(1M_\odot,199 M_\odot)$], so in this region $p(m_1,m_2) = \frac{2}{(198 M_\odot)^2}$.  In 
an $M,q$ coordinate system, in the region consistent with our constraints this prior has the
form $p(M,q) = \frac{4}{(198 M_\odot)^2} \frac{M}{(1+q)^2} $.
The extra factor of 2 arises by compressing the two regions $m_1>m_2$ and $m_1<m_2$ into a single region in the $M,q$ plane (i.e., by requiring $q<1$ or $m_1>m_2$.)

\noindent \textbf{ILEMC.} In the new approach described in this paper, we perform the Monte Carlo procedure described in Eq. (\ref{eq:MonteCarlo})
for a dense and uniform grid in total mass ($M=\lambda_s$).
One-dimensional posterior distributions are found via Eq.(\ref{eq:MCPosterior:1d}).

To be concrete, in Sec.~\ref{sec:Results} our numerical experiment 
will be to
(i) generate a specific list of candidate signals, 
(ii) prepare mock data for the expected LIGO response,
 and finally (iii) apply 
the traditional ILE and ILEMC procedures to 
these synthetic datasets.  We assume both instruments operate at the LIGO O1 sensitivity
\cite{DetectorPaper}. We analyze data segments of 32s in duration sampled at a rate of $16,384$Hz. 

Our candidate signals are nonspinning black hole binaries, with an inclination of $0$ (in the first case) or $\pi/4$
  (in the remaining cases) relative
to the line of sight, and with the distance D scaled so the network signal-to-noise ratio is $25$ or $20$. 
For simplicity, all candidate signals have been generated with a multimodal effective-one-body model
 for nonspinning binary black holes \cite{pan2011inspiral}, henceforth denoted EOBHM.  The EOB surrogate model used in this paper has been trained on EOBHM~\cite{2014PhRvX...4c1006F}, but for signal injection we continue to use the original EOBHM model. The original EOBHM model is also used for parameter estimation with the traditional ILE method while the ILEMC will always use the surrogate model in its analysis.
The likelihood calculation uses frequencies between   $f_{\rm low}=20\unit{Hz}$ and $2000\unit{Hz}$.
A template's 
duration depends on the model.  For EOBHM, the (2,2) mode starts
at $10\unit{Hz}$, to insure  $(4,4)$ mode starts 
before 20 \unit{Hz}.  For the ROM, the entire dimensionless
surrogate model is used when computing Eq.~\eqref{eq:QuantitiesViaSurrogate} so the starting frequency depends on the binary mass; for a sense of scale, at $M=150 M_\odot$ and $q=9$, the
$(4,4)$ mode has a starting frequency of roughly $22 \unit{Hz}$.

\section{Demonstrations}
\label{sec:Results}

Figures \ref{fig:Results} and \ref{fig:Results:2} compare results obtained by traditional ILE and ILEMC applied to identical sources, summarized in Table~\ref{tab:Parameters}, using a range of candidate models that may include or omit higher order modes. Section~\ref{sec:sub:Implementation} provides a complete description of the demonstration's setup.

\begin{table}
\begin{tabular}{l|lllr|c}
ID & $m_1 (M_\odot)$ & $m_2 (M_\odot)$ & $\iota$ & $\rho$ & Model\\ \hline
0  & 35 & 35 & 0 & 25  & EOBHM\\
1  & 100 & 30 & $\pi/4$ & 20  & EOBHM\\
2  & 100 & 50 & $\pi/4$ & 20  & EOBHM\\
\end{tabular}
\caption{\label{tab:Parameters}\textbf{Source parameters}: This table provides the source parameters for each candidate event. All synthetic events were created as distinct realizations of Hanford and Livingston data at a GPS time $10^9$s and a sky
  location of RA=DEC=0. Candidate signals include the $(\ell, m)=$ $(2, \pm 1)$, $(2, \pm 2)$, $(3, \pm 3)$, $(4,
    \pm 4)$ and $(5, \pm 5)$ modes made available by the EOBHM model.
}
\end{table}

\subsection{Single mode, non-spinning EOB surrogate model}

Figure \ref{fig:Results} shows two cumulative posterior distributions from a parameter estimation study performed on a synthetic dataset 
with source parameters (entry ID 0) summarized in Table~\ref{tab:Parameters}
-- a configuration motivated by the parameters of GW150914~\cite{DiscoveryPaper}. 
The solid red curve shows the posterior
distribution 
recovered with traditional ILE and the EOBHM model.  The 
green curve shows the results derived from our new 
ILEMC approach
and a surrogate  trained to reproduce the $(2,2)$ mode of EOBHM. 
Despite the surrogate model ignoring higher harmonic mode content 
we expect these two methods to produce nearly-identical posteriors. Indeed,
this choice of
inclination angle and mass ratio is well known to minimize 
the 
importance of higher modes. For nearly equal-mass binaries
at this relatively low total mass, previous studies have shown 
higher harmonics have 
negligible impact on parameter estimation; see, e.g.,
\cite{2015PhRvD..92b2002G,NRPaper,LIGO-O1-PENR-Systematics,gwastro-mergers-nr-LangeMastersThesis}.

As expected, the 
red (traditional ILE) and green (ILEMC) curves
nearly agree.   
The discrete steps apparent in the ILEMC posterior (green) arise because  of the Monte Carlo  procedure used to estimate the
cumulative; by contrast, the smooth results produced by traditional ILE follow  from 
applying a simple approximation to the function ${\cal L}_{\rm marg,alt}$.  
In \cite{NRPaper}, this approximation was a second-order Taylor series approximation near the peak; in
  \cite{gwastro-PE-AlternativeArchitectures}, this approximation was an interpolating function based on a discrete
  sampling grid.

Here, we have compared parameter estimation (PE) with and without surrogates, using a surrogate tuned to the same model.
Models like EOB have parameters which have been calibrated against numerical relativity in a certain region
  and with a certain accuracy. They have systematic errors, which grow increasingly significant when moving away from the
region where they have been best-calibrated.  For example, even for these binary parameters -- deep inside a
thoroughly-explored region of parameter space --  we do not see this level of agreement when carrying out a similar analysis
using other EOB models \cite{gw-astro-EOBspin-Tarrachini2012}.

\subsection{Multi-mode, non-spinning EOB and NR}

In Figure \ref{fig:Results:2} we show results for  parameter estimation via ILE with EOBHM (solid red); via ILE with
only the leading-order mode from EOBHM (dotted red);  via ILEMC, with a surrogate which reproduces the
$(2,2)$ mode of numerical relativity simulations \cite{gwastro-approx-ROMNR-Blackman2015} (dotted green);  via ILE,
using the same surrogate model (dotted black); and with the same surrogate but including higher-order modes, via ILE
(solid black). The top panels show a
result with $q=0.3$;
the bottom panels show results for  a binary with $q=0.5$. By design,
both scenarios are comparable to  cases examined in \cite{2015PhRvD..92b2002G} with a different parameter estimation
engine (e.g., their Fig 9) using a similar model, albeit at higher SNR.  

\begin{figure}
\vskip20pt
\includegraphics[width=\columnwidth]{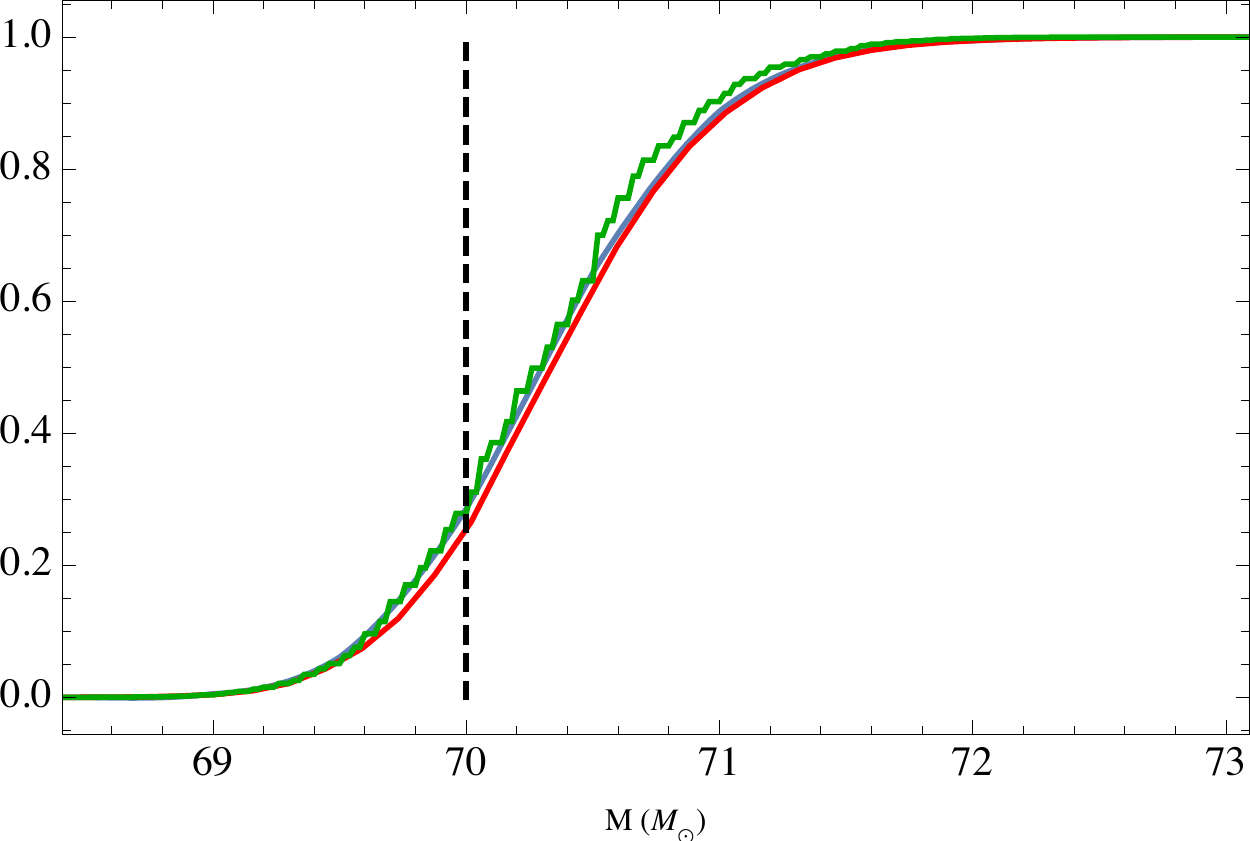}
\includegraphics[width=\columnwidth]{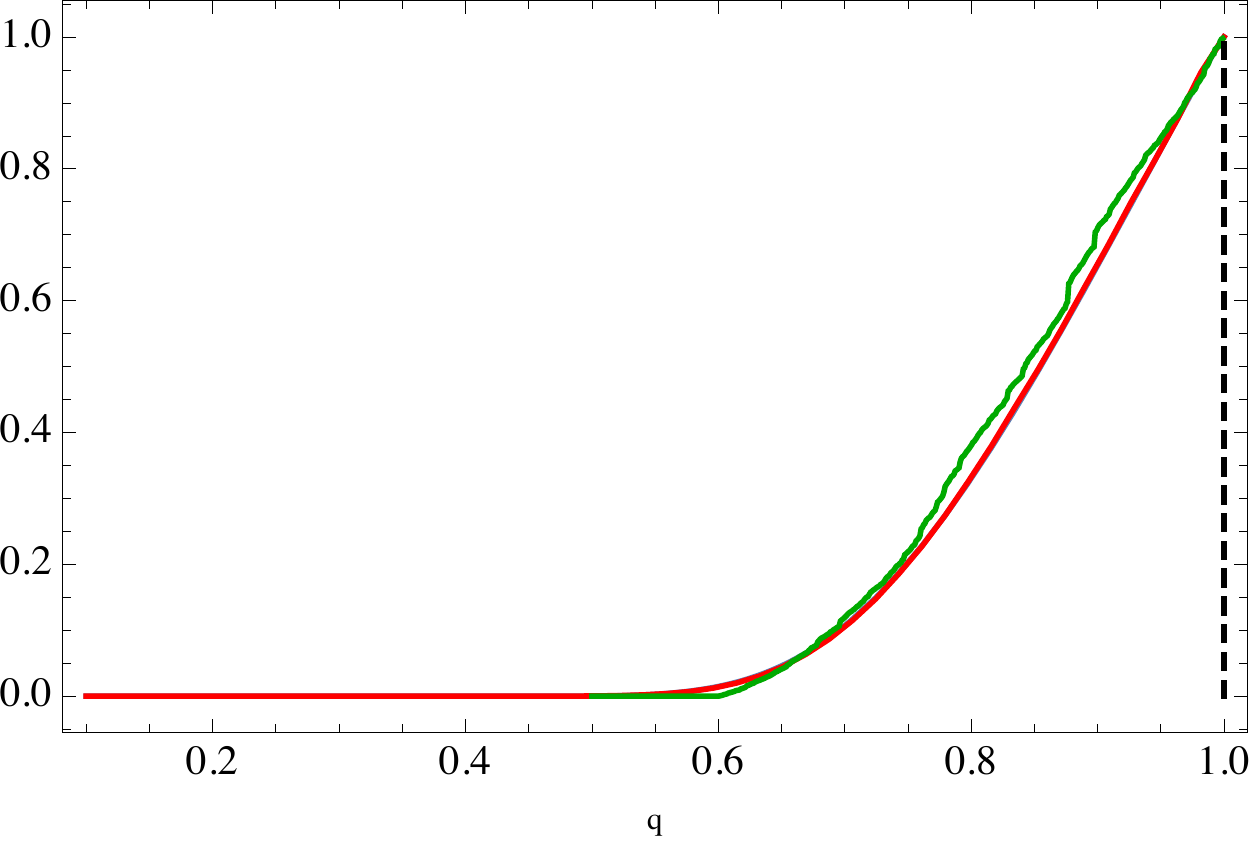}
\caption{\label{fig:Results}
\textbf{Comparison of traditional ILE and ILEMC  pipelines using a GW150914-like signal.} The one-dimensional
cumulative posterior distributions for 
  total mass $M$ (top) and mass ratio $q=m_2/m_1$ (bottom) have been derived
  using both the traditional ILE (thin blue or red curve)
and ILEMC (green curve) frameworks.
The binary parameters
    for this single analysis, summarized in Table \ref{tab:Parameters} (entry ``ID 0"), 
are depicted as a
dashed vertical line. 
Parameter estimation with traditional ILE was performed using the EOBHM model (red and blue) while the ILEMC study uses an EOBHM-calibrated surrogate (green).
To illustrate the consistency between different fitting methods used 
for the traditional ILE results, 
we have considered evaluation of the posterior with a quadratic fit (blue) and a Gaussian process (red).
}
\end{figure}

\begin{figure*}
\includegraphics[width=\columnwidth]{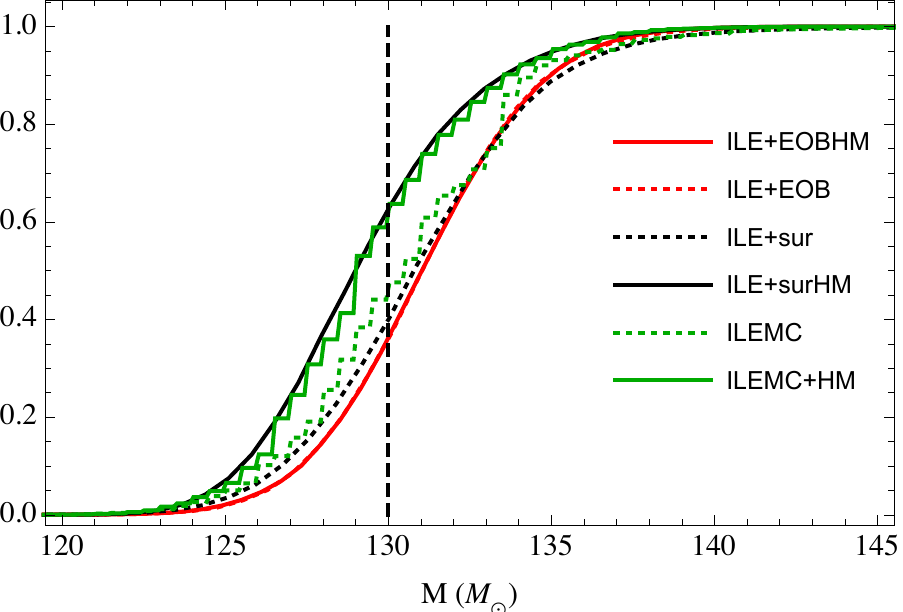}
\includegraphics[width=\columnwidth]{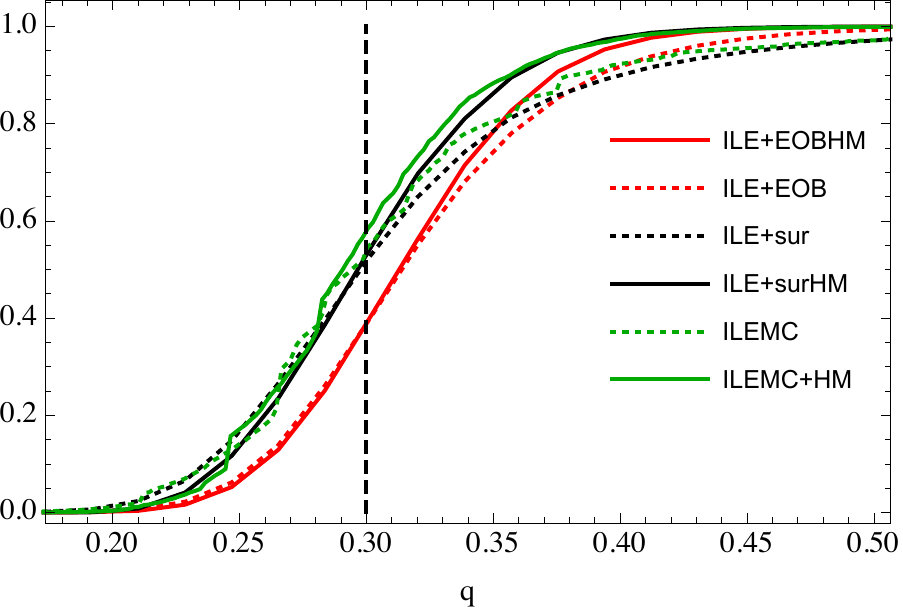}
\includegraphics[width=\columnwidth]{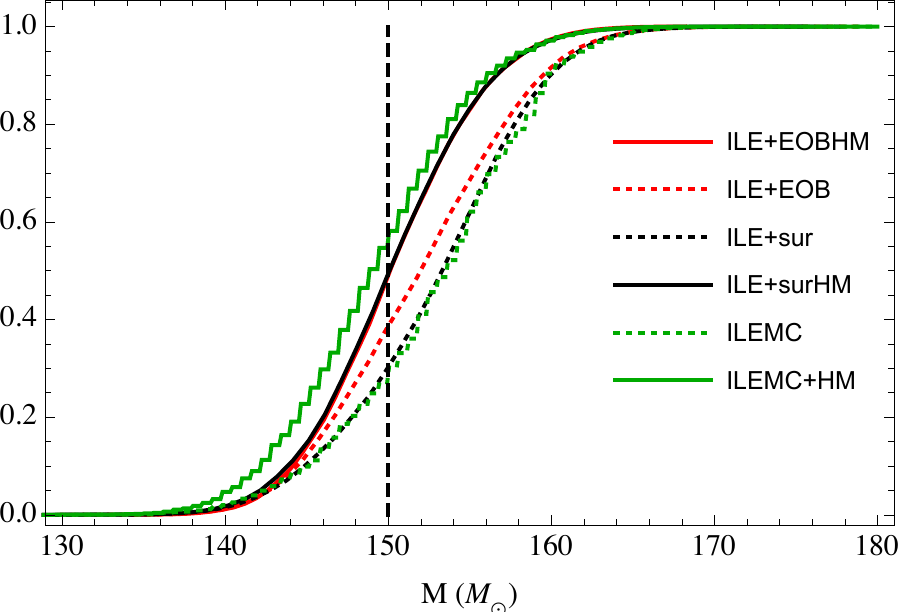}
\includegraphics[width=\columnwidth]{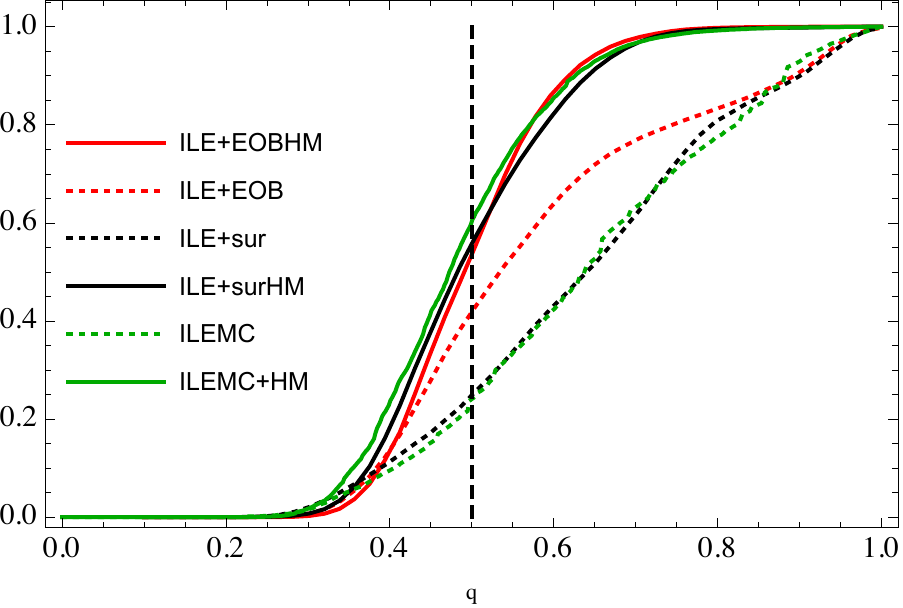}

\caption{\label{fig:Results:2}\textbf{Comparison of traditional ILE and ILEMC pipelines using unequal masses and higher harmonics}: We consider gravitational wave signals
for a source binary with $m_1=100 M_\odot$, $m_2=30M_\odot$ (top panels) or  $m_1=100M_\odot$ and $m_2=50M_\odot$ (bottom panels).
A dashed vertical line depicts these injection parameter values which are also listed in Table~\ref{tab:Parameters}.
Dashed lines indicate results derived using only the $(2,2)$ mode; solid lines indicate results using modes with $\ell \le 4$.
  The red lines indicate PE performed with a nonspinning EOB model and ILE; the green (ILEMC) and black (ILE) lines indicate 
  PE performed with the same surrogate tuned to nonspinning NR simulations \cite{gwastro-approx-ROMNR-Blackman2015} but using two different methods.
}
\end{figure*}

First and foremost, the bottom two panels show that, as with Figure \ref{fig:Results}, ILEMC and ILE agree  when the
models do.
We compare calculations performed using the same NR surrogate model, using ILE (black) and
  ILEMC (green) for parameter inference parameter inference.   For example, in every panel  the dotted black and green lines  -- two independent parameter inference methods using the $(2,\pm 2)$ modes of the NR surrogate -- are in
  good agreement with one another.

Second, the bottom right panel shows by example that at this binary mass higher harmonics matter, since the dotted
and solid lines do not agree.    
As illustrated clearly by directly comparable prior work like \cite{2015PhRvD..92b2002G}, higher harmonics break
degeneracies and improve parameter estimation accuracy.  
The good agreement in the bottom panels of Figure \ref{fig:Results:2} persists despite employing a completely different
model for the source and template.  

Finally, in both the top and bottom panels of Figure  \ref{fig:Results:2}, we see generally good agreement between
  the red curves (PE with ILE, carried out using EOBHM) and the black curves (PE with ILE, carried out using our NR
  surrogate).  Despite systematic differences between the NR surrogate and EOBHM, in most cases this agreement persists
  even when higher modes are omitted (dotted lines) or used (solid lines).

\section{Conclusions}
\label{sec:Conclusions}

In this paper, we unified two strategies to accelerate parameter estimation -- surrogate models  and factored
likelihoods -- and implemented the result in a production-scale environment, ready-to-use on real LIGO data.  
  Our code 
can be used with any (time-domain) linear surrogate model,
 leveraging parallel efforts to better model the multimodal gravitational wave signal from coalescing binaries
\cite{gwastro-PE-AlternativeArchitectures,gwastro-mergers-PE-ReducedOrder-2013,2014PhRvX...4c1006F,gwastro-approx-ROMNR-Blackman2015}.     

With our existing implementation, we can rapidly reconstruct parameters of arbitrary sources
whose runtime is mostly
limited 
by the cost of a low-dimensional adaptive Monte Carlo integral. 
Based on operation counts, we project parameter estimates could be carried out in seconds to minutes.
When operating at its
theoretical limit, this approach can conceivably provide real-time parameter estimation and Bayesian evidence
factors \cite{2009PhRvD..80f3007L,2015arXiv151105955L,2016PhRvD..93b2002K}.  These rapid
calculations will be helpful to improve current detection procedures %
or to supplement investigations into the impact of non-Gaussian noise (``glitches'') on GW parameter
estimation.

Finally, as an illustration of our method's broad utility, we 
have demonstrated
how waveform modeling errors  can, for sufficiently massive 
systems ($f_{low} M > 0.5 \times 10^{-2}$ in dimensionless units), cause us to draw less sharp and reliable conclusions about the progenitor's properties. 
Our method can make direct use of
 high-fidelity surrogate models  trained 
on numerical relativity waveform data without any approximations to general relativity and including all harmonic modes resolved by the simulation. For heavy black holes in particular, where
systematic biases are expected to be the most extreme \cite{2014PhRvD..90l4004V,gwastro-Varma-2016,2016PhRvD..93h4019C}, we demonstrate by example that modeling error 
such as neglecting higher harmonic modes can impact our interpretation of candidate events.

\section{Acknowledgments}
\noindent We acknowledge helpful discussions with
Chad Galley and
Rory Smith, Chad Galley for significant coding effort on the
gwsurrogate project, and both anonymous reviewers for numerous helpful suggestions.
R. O'Shaughnessy was supported by NSF PHY-1505629 and PHY 1607520.  
S.~Field was partially supported by the NSF under award nos.\ TCAN
AST-1333129 and PHY-1606654, and by the Sherman Fairchild Foundation.
The group gratefully acknowledges  Caltech and AEI-Hannover for computational resources.%
\bibliography{paperexport}

\end{document}